# Design constraints for nanometer scale quantum computers.

Ronnie Mainieri

*T-13, MS-B213,*

*Los Alamos National Laboratory,*

*Los Alamos, NM 87545*

ronnie@cnls.lanl.gov

30 November 1993

## Abstract

Nanometer scale electronics present a challenge for the computer architect. These quantum devices have small gain and are difficult to interconnect. I have analyzed current device capabilities and explored two general design requirements for the design of computers: error correction and long range connections. These two principles follow when Turing machines are implemented as integrated circuits. I consider the roles of electromigration through thin wires, circuit layout, and error rates for devices with small gain. The analysis brings into sharp focus the future of nanocomputers and suggests solutions to some of its difficulties. It gives a theoretical model for a nanocomputer, separating the roles of devices and algorithms. Within the model one can implement a stochastic computer, which operates despite quantum device limitations.

## 1  Introduction

Since the construction of the ENIAC, the electronic devices used in computers have shrunk by a factor of $10^5$. The basic driving force for the miniaturization of computers and their components is economic. Integrated circuits, built by lithography, can be produced at very low cost. This economy of scale allows for the construction of computer with a larger number of parts. But the downsizing of the components of a computer cannot go on without limits. Computers are built of condensed matter and the down-sizing has to take that into account.

Before the limit of the atom size is reached there is another limit of importance for the design of computers. It is the point when quantum mechanical effects become important. In the design of a CMOS (complementary metal oxide semiconductor) transistor, quantum mechanics plays a small role. Once the existence of a sea of electrons and holes is given, the transistor can be understood in classical terms. This is because the coherence length of the electrons in silicon is small compared with the size of the transistor. With the use of small gallium-arsenide devices, the coherence length of the electrons becomes comparable to the device size and quantum mechanics starts playing a larger role. At these small scales it is no longer true that the device can be treated as an isolated unit, that any amount of gain is possible, or that any amount of fan-out can be achieved. The nanodevice is not like the CMOS transistor. Does this have any implications for the design of computers?

I want to examine the implications of quantum devices in the design a modern electronic computer. What concepts are essential for the operation of a computer? Can these concepts be implemented with nano-scale electronics? To answer these questions I will draw from several areas: VLSI (very large scale integration) design, estimates of errors, and the physics of electron conduction through semiconductors. I will limit myself to semiconductors not because of any fundamental reason, but for economical reasons. Computers are constructed using VLSI integrated circuits in factories that cost close to $10^9$ dollars and take five years to become operational [1]. This is at the limit of most commercial enterprises [2].

The results also apply to quantum computers. Most research on quantum computers is done in the abstract [3]. A Hamiltonian is proposed and it is shown to have certain interesting properties as a computer. It is implicit in the research that a realization of the Hamiltonian can be found in condensed matter. Most designs emphasize condensed matter because the interactions are stronger between electrons than they are between photons. I will not be investigating the possibilities that physical laws offer as algorithms.

For the analysis of a computer I will consider it as a central processor that is connected to memory. The processor has to keep its state from one clock cycle to the next (a finite state automaton). It can be implemented as a Boolean function where some of the output can be used as input in the next clock cycle. This is discussed in section 2. A typical Boolean function of $n$ bits is realized as a circuit with less than $2^n/n$ gates, and it is typical for it to have more than $n$ gates. If we use wires to connect the output of gates to the input of others, then the typical number of wires fanning-out from each gate will be larger than $n^{1/2}$. This means that the layout of the gates in two dimensions will not be planar. These estimates are confirmed experimentally in the form of Rent's law for the number of connections.

Thin wires (say, less than 250 nm) are difficult to build in integrated circuits. They are usually destroyed by the currents used in intergrated circuits. As the electrons flow through the wires they collide with the atoms in the grains that make them up. This creates a slow diffusion of atoms in the direction of the current flow. After some time the vacancies that the atoms left downstream will accumulate close to the contact of the wire and break open the contact. This is the process of electromigration. It is discussed in section 3.

The devices in an integrated circuit are not ideal. There are manufacturing imperfections. There are temperature fluctuations. Any realistic design must take these variations into account and error correct. Error correction can be done by digital means or by analog means. I will show that digital error correction cannot overcome the large error rates of nanodevices in section 5. In current digital computers errors are corrected by using gain — an analog method. For transistors the gain curve is nonlinear. This allows variations from both the 0 bit or 1 bit voltage to be corrected to their standard value. This observation, and its importance for computer design, was first published by Keyes [4]. Analog error correction is a statistical effect arising from the sea of electrons in the semiconductor. It depends on the large size of the device to keep statistical fluctuations small.

The problems with wires and errors would seem to indicate that it is impossible to build a computer with nano-scale electronic devices. But this is not the conclusion from the argument. What seems impossible is a general purpose computer with the current low gain devices. If a device is invented that has larger gain and the lifetime of thin wires can be extended to several years it should be possible to build a general purpose computer. Even if these goals are not achieved, it should be possible to build a useful computational device. There are many algorithms that operate despite errors in the state of the computer. In section 6 I will introduce an example of a stochastic computer and a class of algorithms it may execute.

## 2  Boolean graphs

In this section I am going to argue that computers need long wires. The basis of the argument is in graph theory. A Boolean circuit, which exists in any processor, can be represented by a graph. Which graph will depend on the exact computer, but I will then show that graphs from Boolean circuits are complicated. For the simple case of planar graphs it is possible to estimate how long the wires in the computer have to be. But planar graphs are too simple for a computer. That is because the most complex language they can recognize are regular expressions, and most computing tasks cannot be represented as regular expression. For more complicated languages, the wires have to be even longer.

The function of wire is to carry a bit. There are several ways to accomplish this in a circuit. The most common way is to apply a voltage to a wire made of aluminum, copper, gold, or silver. Metal wires have the great advantage of speed. In principle they can carry a bit at the speed of light; in practice they are limited by the time it takes to charge up the metal wire. That time should be shorter than the clock speed and is of the order of RC, where R is the resistance of the wire, and C its capacitance. But metal wires are not the only way to carry a signal from one part to another in a circuit. For example, the bit can be actively carried through the circuit by an active wire. An active wire is a chain of devices that pass a bit unaltered from one end to another. The transfer is synchronous with the clock (or cycles) of the circuit. Therefore an active wire is not as fast as a copper wire. In billiard ball models of computers [5] the signals are carried by billiard balls and the wires are the free flight regions of the billiard. The free flight regions act as active wires, because it takes the signal a certain number of clock cycles to go from one part of the circuit to another. Active wires slow down a computer: they limit how fast an algorithm may execute.

There is a rule that relates the number of wires in a circuit to the number of gates. It is Rent's rule [6]. It is an empirical power law relating the number of interconnections (or wires) $w$ to the number of gates $g$:

$$w = w_0 g^r .$$

The exponent $r$ is the Rent exponent and varies with the type of circuit. Assume, for example, that the circuit is a memory chip. For each storage unit (flip-flop or capacitor) there is a fixed number of wires. To read and write the bits there is an array of wires forming a grid. There are also power supply and ground wires; these also form a grid. The wires in a memory chip are just a series of grids superimposed. The number of wires in the grid is proportional to the perimeter of the memory. The number of storage bits (gates) is proportional to the area of the memory. This implies that the number of wires is proportional to the square root of the number of bits. The constant $w_0$ will depend on the exact number of grids that are required for the memory to operate. But independent of the value of $w_0$, it shows that the Rent exponent r should be 1/2 for a memory device. This is confirmed in experimental measurements [6].

For circuits that are not as regular as a memory, the Rent exponent is larger than 1/2. This means that there more interconnections among the circuits of a processor than there are in memory device. Later I will argue that processors execute Boolean functions that are more complicated than memory lookup.

Being more complicated they have to transmit more information through the circuit than a memory does.

It may be possible to reduce the number of connections in a central processor. In traditional designs the number of gates is minimized and the Rent exponent is larger than 1/2. But it may be possible to design central processor where the communication time is minimized, leading to smaller Rent exponents [7].

## 2.1 Boolean functions as graphs

A model of a computer is a large amount of memory connected to a processor. The processor is a finite state machine. It looks at the content of the memory and changes its state accordingly. This change of state may change the memory or not. The finite state machine executes a Boolean function. It is made from a series of interconnected AND, OR, and NOT gates.

To study the types of connections and number of wires in a computer it useful to simplify it. The exact type of gates or memory units that are being connected do not really matter for the analysis of the number of wires and their lengths. I will then replace each gate of the computers by a node. The circuit reduces to a set nodes interconnected by wires. It is a graph in the sense of graph theory. In this model of an electronic circuit all the inputs of a gate converge to a single node or vertex The wires that interconnect the gates form the edges of the graph.

A graph is a set of nodes or vertices connected by edges. The positions of the nodes or how they are embedded in space does not matter in the definition of a graph. All that matters is that there are nodes, and that there is a pairing of nodes in the form of edges. Two nodes of a graph are connected to each other if there is a sequence of edges that takes you from one node to the other. If every node of the graph is connected to every other node, then the graph is connected. It is common to think of graphs as embedded in the plane. Then nodes become points and the edges become lines. If the graph can be drawn in the plane so that no two edges (lines) cross, the graph is planar. In general graphs that are too interconnected cannot be planar.

A simple example of a circuit and its graph is shown in figure 1. In the circuit, figure 1(a), there are four wires comming in from the left, the inputs. There is one output comming out from the right. There are a total of four gates in the circuit. Each gate of the circuit gets replaced by a node and all the inputs of the gate end up coming into the same node. The circuit in figure 1(a) becomes the graph in figure 1(b).

One can already see the difficulty in laying-out circuits. As each gate occupies some area in the integrated circuit, lets assume, as this only changes estimates by a constant, that all gates can be placed on the nodes of a square grid. If a given node has many lines connected to it, then some of the nodes on the other end of these lines will have to be far away. For example, if the nodes are limited to be on a grid of unit size squares, then a given node will have 4 neighbors at distance one, 4 at distance $\sqrt{2}$, 4 at distance 2 and so on. So if the central node had to be connected to 12 other nodes, there would have to be a wire of at least length 2.

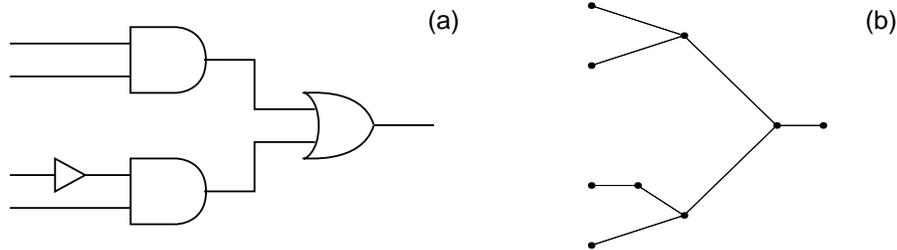

Figure 1: The graph equivalent of a circuit. To construct a graph of a circuit (b), every gate of an electronic circuit (a) is replaced by a node.

So the more nodes are interconnected, the longer the wires of the circuit have to be. How long depends on how efficiently gates can be arranged on a square grid and how how interconnected each node is. In section 2.2 I will explain how the simplest useful task a computer can do — recognizing a regular expression — requires a circuit corresponding to a planar graph. Laying-out a planar graph requires laying-out the tree graph it contains 2.3. From the tree graph on can estimate a lower bound on the longest wire needed in a circuit. Other circuits 2.4 are even more complicated and require more gates, making the wires longer as in Rent's law.

## 2.2 Regular expressions

The simplest circuits are those that recognize regular expressions. A regular expression is a simple language that is used to represent all the possible patterns that may occur within a context. Regular expressions are equivalent to finite state automata, the processing element that occurs in the Turing machine. An example of the use of regular expressions is in describing an identifier in the C programming language. An identifier is a sequence of characters. The first character must be a letter or the underscore. The remaining may be letters, numbers, or the underscore. For those that are familiar with UNIX[1] regular expressions, the C language identifier could be recognized by

$$\mathtt{[a\text{-}zA\text{-}Z\_][A\text{-}Za\text{-}z0\text{-}9\_]*}$$

The ranges within square brackets mean that one of those symbols should occur. And the * means that the previous character (or range of characters) may occur zero or more times. The importance of regular expressions in the design of processors is that sequence of events can usually be specified by regular expressions. Usually, and not always, because there may be cases in which one has to compare a large number of events before deciding to take action. Situations where one has to recognize

$$a^n b c^n,$$

and n is not known in advance, are not well suited for regular expressions.

Regular expression can be built with planar graphs (as shown by Mukhopadhyay [8]). Floyd and Ullman [9] have given a practical algorithm that translates a regular expression into a circuit description. To explain their algorithm I would have to introduce many related concepts, so I will just indicate a textbook (reference [10, sec. 3.4]) where the layout algorithm for regular expressions is discussed. An important consequence of being a planar circuit is that the number of gates needed to recognize a regular expression with n symbols is proportional to n.

## 2.3 Trees

To set a lower bound on how long wires have to be, I will consider a special subset of graphs: trees. Trees play a special role in graph theory. Every graph, no matter how simple or complex, has as a subset a tree. I will first give an informal definition of trees. Then I will show that there is a very efficient scheme to layout trees in the plane. This scheme will give a lower bound on the length of the wires that are required to implement the circuit that the tree represents. Because trees are subsets of all graphs, circuits that are more complicated require wires at least as long as those of the tree.

Trees are graphs that do not have loops of edges. In a tree, starting from any node it is not possible walk along the edges and return to the starting node without repeating an edge. If we start with a complicated graph and remove enough edges it can be reduced to a tree. So trees are simple graphs, and complicated graphs contain them. There are many possible trees, but any particular tree has a maximum degree. The degree of a node is the number of edges connected to that node. Assume that the maximum degree is three. This is the case when every gate has one or two input terminals and only one output terminal. The complete binary tree would be the case when every node has exactly two input terminals and one output terminal. Figure 2 has three examples of trees. In tree (a) all nodes have degree three or less. Therefore it can be embedded into a complete binary tree, which is shown in (b). Tree (c) is an example of a tree that is not binary.

---

[1] UNIX is a registered trademark

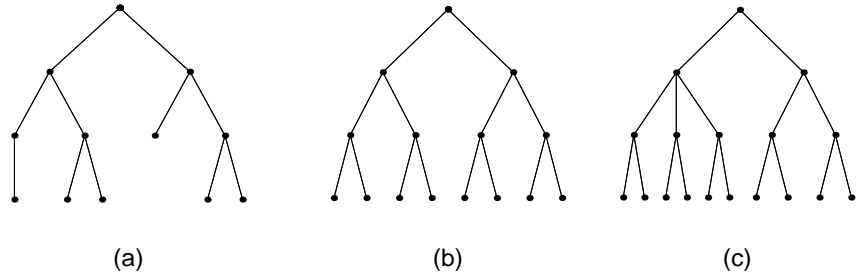

Figure 2: Trees with degree smaller that 3 (a) can be embedded in a complete binary tree (b). If the tree has a degree 4, as (c), then the embedding must be done on a complete ternary tree.

The most efficient way to layout a tree in a grid is using an H-tree. The H-tree arrangement uses the least area. It is called H-tree because it is formed from a series of nested structures that resemble the letter H. The construction is recursive. In figure 3(a) I show the first step of the construction and in (b) a larger example. To compute the area of this tree, notice that each node occupies a finite amount of area, 4 unit squares in the figure. If the tree has $n$ nodes, then the area occupied will be proportional to $n$. The longest wire used in the tree is the one connecting the root node to one of its leaf nodes. The side of the square containing the H-tree has length $\sqrt{n}$, and the longest wire has length of $\sqrt{n}/4$. It turns out that it is possible to layout the complete binary tree so as to minimize wire length. To determine the longest wire length one notices that in the H-tree arrangement the length of the longest wire doubles each two new branch-levels the tree gains. This leads to a wire of order $\sqrt{n}/\log_2 n$ (see [10, sec. 3.2]).

## 2.4 General Boolean functions

General purpose computers need to process languages that are more complex than regular expressions. It has to recognize (or generate) patterns from other languages (see table 1). But these more complicated languages have regular expressions as a subset, and therefore the graphs and circuits for these languages will be as complicated as those for regular expressions. This means that the bound on wires of length $\sqrt{n}/\log_2 n$ must hold as a lower bound. The $n$ is the number of nodes in the circuit. The exact relation between the number of gates in a circuit and the language it recognizes is still an active area of research. There are no general theorems relating the two.

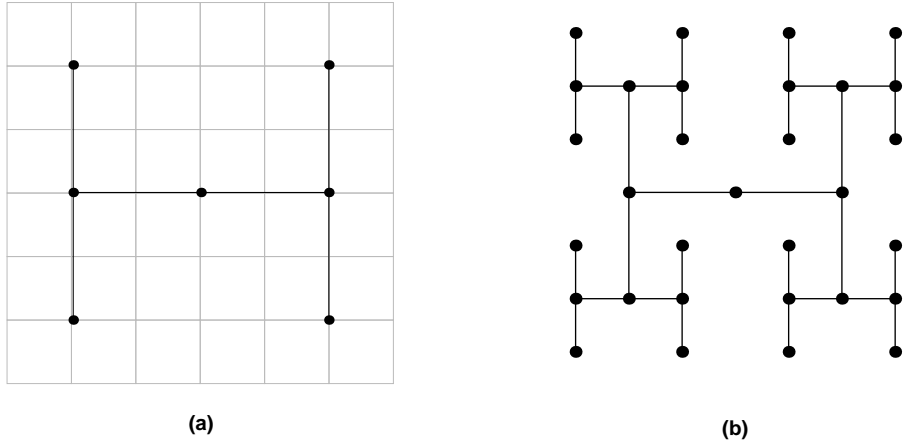

Figure 3: The H-tree is the most area efficient way to layout a complete binary tree. The construction process is recursive. The first step is in (a) and is laid over a grid. Each node occupies four unit squares. Part (b) shows a more complicated example.

A less general approach is to ask what is the relation between the number of gates and the number of inputs to a Boolean function. The first to give a satisfactory answer to that question was Shannon [11]. He showed that if you devised an algorithm that transformed a Boolean function of n inputs into a circuit, then you would need at most $2^n/n$ gates to implement the function. He also showed that there is a function that requires that many gates. Does that mean that the typical Boolean circuit requires an exponential number of gates? The answer is yes. A typical Boolean function of n inputs and one output requires $2^{n-3}/n$ gates [12].

| Language | machine |
|---|---|
| Regular expressions | finite automaton |
| Context free | stack automaton |
| Context sensitive | bounded automaton |
| Recursively enumerable | Turing machine |

Table 1: Different types of languages and the computer that recognize them. Regular expressions are the simplest requiring a finite automaton, and recursively enumerable languages are the most complex requiring a Turing machine.

In practice the circuits that are used in processors are not as complicated. Multipliers, sorters, control units, all have power law dependence on the number of inputs. That is because these circuits are designed by people, that often use structured methods to design the circuits. The gates are combined in modules; the modules into larger circuits. Each modules execute a simple function, such as storing a byte, adding two small numbers, or comparing two small numbers. More complicated functions are programmed in the circuit and executed in many time steps, therefore trading circuit complexity for time of execution. The modular structure tends to diminish the number of gates and also diminish the number of interconnections among parts of the circuit. From the Shannon estimate for the number of gates in a circuit one can see why should the circuit be simpler when it is built out of modules. Let me start with a simple example. Assume that a function of 2n bits, $f(\epsilon_1, \ldots, \epsilon_{2n})$ can be computed by the logical AND of two smaller functions $g_1$ and $g_2$

$$f(\epsilon_1, \ldots, \epsilon_{2n}) = g_1(\epsilon_1, \ldots, \epsilon_n) \wedge g_2(\epsilon_{n+1}, \ldots, \epsilon_{2n}) \ .$$

The smaller functions $g_1$ and $g_2$ represent the modules of the circuit and the larger function f represents the whole circuit. From Shannon's estimate, the function f should require $2^{2n}/(2n)$ gates. The smaller functions require $2^n/n$ gates each. Because f is the AND of the smaller functions, it then only needs $1 + 2^{n+1}/n$ gates. A number that is the 1/2 power of the initial estimate for the number of gates. The notion of functional composition can be generalized and it rapidly brings down the number of gates needed.

Shannon observed that the majority of Boolean functions cannot be written as the composition of smaller functions. So in general it is not possible to bring down the number of gates that compute some arbitrary Boolean function.

From Shannon's result the number of gates in a circuit with n inputs will have $2^n/n$ gates. To layout these gates will require a circuit with wires as long as $\sqrt{2^n/n}$. Even if we limit ourselves to circuits that grow as $n^a$, for some small exponent a, the length of the wires will be proportional to $n^{a/2}$. This means that for devices on a grid there will always be gates that have to be connected to a large number of other gates, all of which cannot be nearby. Long wires cannot be avoided.

## 3 Physical wires

Wires carry signals within an integrated circuit. They are made of metals and they are the last few layers in the fabrication process of an integrated circuit [13]. The resistance of a wire depends on the material used, and at the nanoscopic scale, on its shape (see the basic theory of Landauer [14] and Büttiker [15] and the results of Roukes and Alerhand [16]). The capacitance and inductance of the wire come from stray couplings to the rest of the circuit. Just as a transmission line, a wire can be substituted by resistors, capacitors, and inductors. For the wire shown in figure 4(a) the equivalent circuit (figure 4(b)) is the transmission line with total resistance R, total capacitance C, and total inductance L. A current j of frequency $\omega$ along this wire is given by

$$j(x,t) = \text{Re}\big(e^{i\omega t}\big(j_a e^{\gamma x} + j_b e^{-\gamma x}\big)\big) \ .$$

The equation for the voltage would have a similar form. The attenuation of the signal as it goes over the wire is given by $\gamma$

$$\gamma = \sqrt{i\omega RC - \omega^2 LC} \ .$$

This signal propagates along the wire with speed $-i\omega/\gamma$. If the wire is long (much larger than $1/\big(\omega\sqrt{LC}\big)$) then it can be treated as a transmission line. Signals of different frequencies will propagate at different velocities with small attenuation. A wire in the transmission-line mode can be used to send high frequency signals. If the wire is longer than the attenuation length (much larger than $1/\big(R\sqrt{C/(4L)}\big)$) then the wire can be treated as an RC circuit. If a signal is injected into a wire, then the other end will receive an exponentially attenuated signal at first. This signal will propagate with velocity $-i\omega/\gamma$, as in the transmission-line mode. Then, it will grow to its maximum value in a time RC.

For computers that operate under the gigahertz frequency, the velocity of light is not a limitation in circuit design. What sets limits on the operating frequencies is the time it takes wires to charge up. The speed of propagation of the first wave of electrons is always $-i\omega/\gamma$, but the time it takes to reach a certain voltage depends on RC. When a signal is injected at one end of a wire the voltage at the other end grows as $j_0(1 - e^{-t/RC})$, with $j_0$ the current of the signal. If a larger current is put through the wire, larger voltages can be obtained at the other end in shorter periods. But the current through a thin wire cannot be increased beyond a limit.

Let me explain the limitation. The image of a metal wire is a crystal that conducts electrons, with maybe a few point defects and dislocations spread around. That is true, but only at very small scales (a few nanometers or less). A typical wire is a polycrystal [17]. It is formed from many crystal grains that have different orientations — a polycrystal. The interface between the grains occupies a larger volume than the typical defects of crystal (vacancies and dislocations). And it dominates the transport properties of the metal wire at low temperatures. For example, at room temperature the movement of atoms along the grain boundaries can be $10^6$ faster than through the lattice.

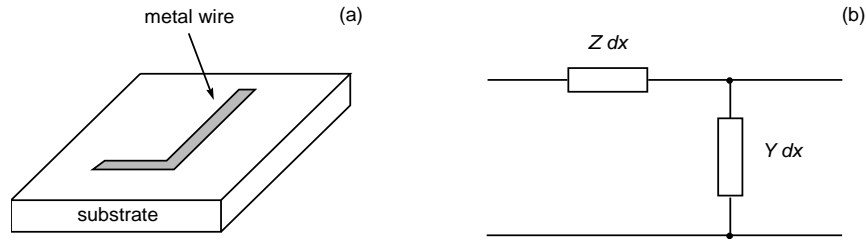

Figure 4: Wire on an integrated circuit (a). It has a resistance R which depends on its shape, and capacitance C, and inductance L, which depends on its coupling to the rest of the circuit. The wire is equivalent to a circuit with impedance Z and admittance Y per unit length (b).

Electromigration is the movement of the atoms of a wire as electrons flow through the wire. As the electrons flow from the positive to the negative terminal of a wire they create an electron wind. As they blow past the atoms of the metal, they collide with them, moving them upstream and leaving vacancies behind. The vacancies eventually get filled in by other atoms that are moving upstream. The net effect is that the atoms move upstream and the vacancies move downstream. Eventually they reach the end of the wire where the contacts are. Near the negative contact the vacancies pile up forming a large hole. This hole diminishes the available cross section for the electron wind, increasing the current density. Larger currents mean higher temperatures and accelerated electromigration. The hole near the negative contact grows as the current continues to flow, and when it becomes as large as the wire, the wire breaks open. On the positive contact the atoms start piling up. This forms a hillock which spills out from the wire and creates a short circuit. If the vacancies and atoms go beyond the ends of the wire, they will form a spike through one of the layers of the device and again lead to a failure. Too much electromigration will destroy a wire.

How long will a wire last? As wires becomes sub-micron in width their lifetimes shorten due to electromigration breakdowns. The median time-of-failure, $t_{50}$, gives the lifetime of a wire. It is the time it takes for half of the wires in a batch to fail. The lifetime of wires due to electromigration is very difficult to characterize and many factors seem to contribute to it: the size and stacking order of the grains in the polycrystal, width and length, and number of bends in the wire. The most important factors are the operating temperature T, the current j and how difficult (in terms of activation energy $E_a$) it is to move an atom. These three factors can be combined in an Arrhenius-type empirical formula

$$t_{50} = \frac{A}{j^n} \exp(-\frac{E_a}{k_B T}) \;,$$

where $k_B$ is the Boltzmann constant and n is an empirical exponent that varies between 1 and 15. Most theoretical models of electromigration predict 1 or 2 for the exponent n. The constant A has to absorb all the other factors that affect the lifetime of the wire. The effects of a thinner wire are contained in the constant, so it is only through experiments that the lifetimes of thin wires can be determined. The constant A is supposed to be proportional to the cross section and to depend on the length L through $e^{1/L}$. A series of experiments with thin wires in intergrated circuits was carried out by Kwok and collaborators [18]. They find that wires of sub-micron size will have lifetimes of the order of $10^2$ hours under realistic working conditions. This is a much shorter lifetime than needed for a useful integrated circuit.

I will take the short lifetime of thin wires as an indication that a nanocomputer cannot have long wires.

## 4  Small devices

There are a large number of devices in the literature that could serve as a basis for a nano-scale computer (see the reviews in reference [19]). To understand the necessity for error correction, I will abstract one property of these devices: that they use few electrons to operate. I see three different modes of operation for small devices: quantum interference, resonant tunneling, and bound-state. In quantum interference and resonant tunneling devices charge is transported through the device. In bound-state devices charge can be put into and taken out of the device, or charge may be excited into a different state.

Most of the nano-scale devices operate at liquid nitrogen or liquid helium temperatures. Because of their size, the voltages used to operate them are small, and the thermal energy much be kept low. The plot in figure 5 delimits the range of operations of many of the nano-scale devices. For comparison, CMOS transistors are marked of the diagram.

A quantum interference device operates by splitting the wave function of an electron along two different paths and later joining it. Along one of the paths the phase of the wave function is changed. If it is changed by the right amount, when the path reunite there can be destructive interference. A quantum interference device is shown in figure 6. The phase change can be controlled by an electric or a magnetic field. For a quantum interference device to operate, it must be precisely dimensioned. It can neither be too long nor too short if the interference effect is to work as a switch. It must also operate at very low temperatures to avoid thermal gradients and phonon-electron scattering that could destroy phase coherence.

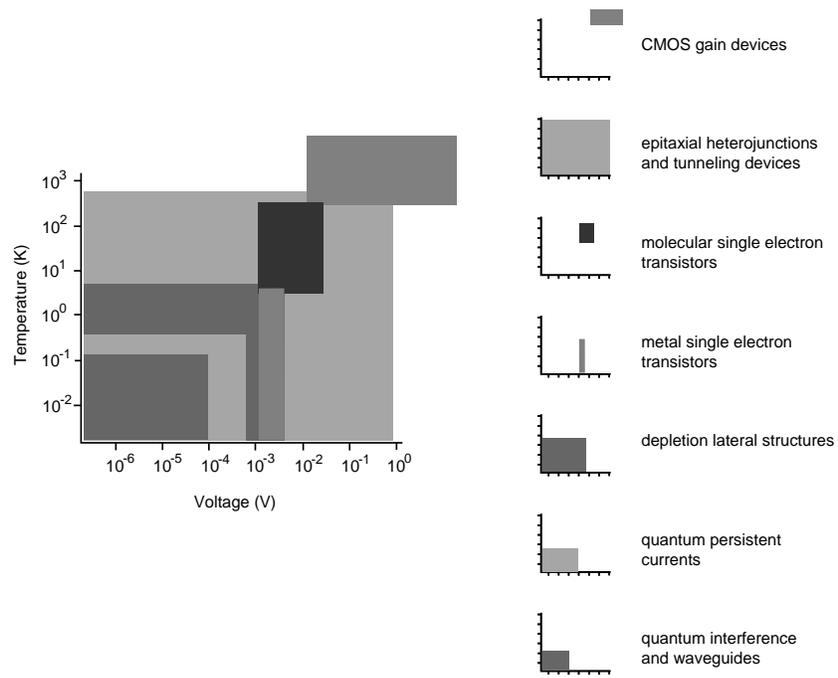

Figure 5: Parameter range for quantum effects in nano-scale devices. The typical range of operation for CMOS devices is in the upper right hand corner of the diagram. (Adapted from a diagram of Mark Reed by permission.)

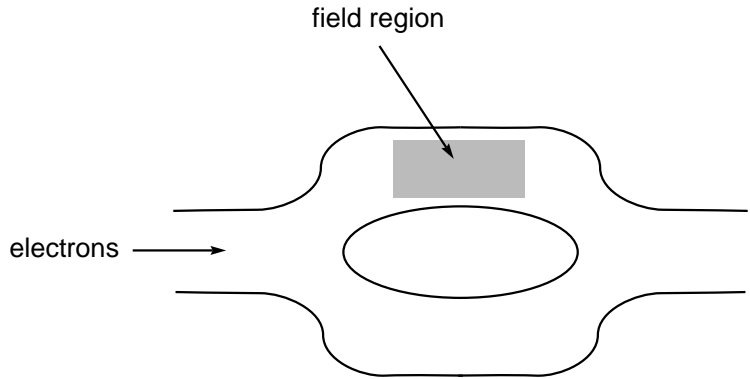

Figure 6: A quantum interference device. Electrons are injected through the left, the input, and exit through the right. There are two possible paths for the electrons. Through the upper branch the phase of the wave function can be changed with the external field.

Resonant tunneling devices operate by controlling the tunneling rate between potential barriers. The typical arrangement is shown in figure 7. There is a double barrier between the Fermi sea at the left and the lower levels at the right. If there were no barrier the electrons would flow from the left to the right. The double barrier creates a one-dimensional well. The well will have at least one bound state, and often more. If the levels of the well coincide with the energy of the electrons on the left, the tunneling rate through the double barrier is enhanced. If the levels do not coincide, the tunneling rate is reduced. This is the basic idea for a series of devices. It was invented by Esaki in the late 1950's [20, 21]. The exact form in which the position of the levels of the barrier are controlled, and the number of barriers leads to different devices. Examples are the Coulomb blockade devices, resonant tunneling diodes and transistors, and single electron transistor. All these devices share the same problems: for their operation two different energy levels must be precisely matched: those of the sea (left in figure 7) and those in the discrete levels of the well.

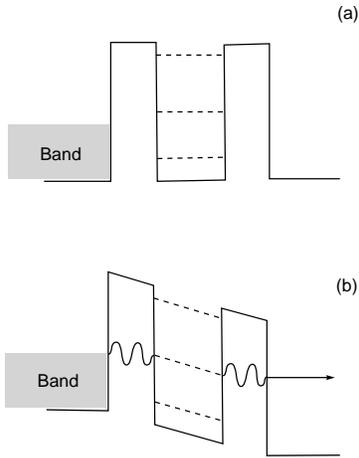

Figure 7: The resonant tunneling diode. If the Fermi energy does not match one of the levels of the well, the current through the device is reduced. By applying an external bias the levels of the well can be made to match the Fermi level, enhancing the current.

Bound state devices use discrete quantization of the momentum of an electron in one or more directions. An example is the quantum well [22]. In the quantum well the momentum of the electron is quantized in the vertical direction, but in the other two directions the levels are so closely spaced that they can be considered a continuum. The net result of this geometry is that electron bounces within the well as if it were restricted to a two dimensional surface. Bound state devices can be used to store information. The amount of charge in a well, or state of a large molecule can be used to store the bit. The reliability of these devices depends on the number of particles (electrons) involved in the device. If a large number of electrons is used, then the loss of a few electrons to the substrate should not affect the performance of the device. If a small number of electrons is used, a loss of a few can change the value of the bit stored.

The devices I have examined share the property that only a few electrons are involved in their operation. With few electrons it is difficult to have large gain. Gain for small devices is the transfer of electrons from a reservoir to the wires of the circuit. When the controlling fields of the device involve just a few electrons it becomes difficult to isolate and control the flux of electrons from the reservoir to the wires of the circuit. Also, a small number of electrons makes for poor statistics in defining the logic states. With small gain, errors become more likely and there is less room for variation in the parameters of the device during fabrication.

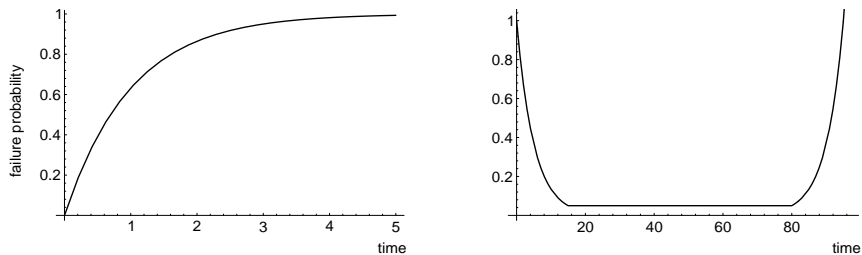

Figure 8: Exponential and bathtub failure rates for devices.

## 5 Errors

All computers fail at some point. When a component of a von Neumann computer fails, the computer fails. A von Neumann computer is a Turing machine, and Turing machines compute the value of recursively defined Boolean functions. If one of the bits of the computer changes (from a zero to a one or from a one to a zero), then the function being computed is different and the result of interest may be wrong. In most cases the computer will crash. The time the computer can go without a crash sets the limits on what calculations can be done with it. It limits the number of bits that may be stored. And it limits the number of cycles the computer may execute without errors.

A device may fail for two reasons: it breaks, or it makes an error. When a device breaks it has to be replaced by a working one. When a device makes an error the program the computer is executing produces a wrong result. Transistors (and therefore logic gates) break before they make an error. Dynamic memories (DRAM) use a leaky capacitor for storage and make an error before they break. The two failure mechanisms are statistically different (see figure 8). The transistor devices have a failure rate which follows the "bathtub" distribution [23, sec. 2.26]. This means that the probability of failure, after the initial burnout, is very small and increases only after the lifetime of the transistor has expired. The lifetime can be of the order of years. The memory capacitors have a failure rate which follows the exponential distribution. The probability that they will fail increases exponentially with time. If the state is kept too long in the memory without refreshing, the memory fails.

Transistors break before they make an error because their principle of operation involves large number of electrons (of the order of $10^{18}/\text{cm}^3$). The large numbers makes the transistor a reliable gain device. A transistor failing to amplify is like all the air of a room going to one side — not impossible, but very unlikely. Devices that operate with just a few-electrons are not as reliable. The event of an electron failing to tunnel through a barrier, or being knocked off into another semiconductor layer is not small.

Their are many mechanisms by a which a device fails [23]. I would like to concentrate on few-electron devices, which make an error before they break. To keep the discussion general I will idealized the digital computer built from few-electron devices as being a set of devices. A device can be a logic gate or a memory unit. When I say a computer, I mean the processing units and memory units. (For nanocomputers it may be that processing and memory are intermixed.) At first I will disregard the interconnections. Each device will have a probability $\epsilon$ of failing in one time step of the computer. The probability of one device failing is independent of the state of other devices, and for how long the device has been operating. When a device fails, it has at its output the wrong bit. If it is a logic gate, then it has the wrong output; if it is a memory device it is storing the wrong bit. This probability can be thought of as a small number, although in some cases in may be as large as 0.01. As with any von Neumann computer, if one device fails, the computer fails.

## 5.1 Failure rate for unreliable computers

Their are two questions to ask about this unreliable computer: how many devices can it have? and how many time steps can it go? Assume that the computer has B devices. The probability that one or more have failed is given by the binomial distribution. If exactly one device failed, it could be any of the B devices. The failure could happen in B ways. If exactly two devices failed, then the failure could happen in $B(B-1)/2$ ways. The probability of failure $p_f$ for the computer is the sum of all the ways it could fail,

$$p_f = \sum_{1 \leq k \leq B} \binom{B}{k} \epsilon^k (1-\epsilon)^{B-k} .$$

If $\epsilon$ is a small number, then the probability of failure $p_f$ can be approximated by

$$p_f = B\epsilon + O(\epsilon^2) .$$

If one of the devices fail, the computer fails. The number of devices that have failed is $Bp_f$, so for a working computer we must have that

$$Bp_f \leq 1 \text{ or } B\epsilon^2 \leq 1 ,$$

which sets an upper limit $B_{max}$ of the number of devices in the computer to

$$B_{max} = \frac{1}{\sqrt{\epsilon}} .$$

This simple estimate shows that one has to work very hard on the reliability of the components to have a computer with many devices. Later we will see that error correction allows the maximum number of devices $B_{max}$ to be increased with a fixed error rate $\epsilon$.

A similar calculation shows how many time steps the computer may execute before it fails. Assume that a computer goes for a total of C cycles (time steps). We want to compute the probability $p_C$ that it will fail after C time steps. This means that the computer worked for at least C steps. It may fail at the $C + 1$ step or it may fail at a much later time. If $w$ is the probability that it worked in one time step, then

$$\begin{aligned} p_C &= \sum_{k \geq C} \text{worked k cycles, failed on k} + 1 \\ &= \sum_{k \geq C} w^k(1-w) \\ &= w^C(1 + w + w^2 + \cdots)(1-w) \\ &= w^C . \end{aligned}$$

We want to determine the probability that it fails. Let us assume that it happens when the probability of working is $1/2$. The value of $C_{max}$ for that to happen is $\ln(1/2)/\ln w$. If we had chosen a fraction different from $1/2$, it would just change the coefficient multiplying $1/\ln w$.

The two results for failure can be combined. As the probability of a computer with B devices to work is $w = (1 - B\epsilon)$, the maximum number of time steps it may execute is of the order of

$$C = -\frac{\ln 2}{\ln(1 - B\epsilon)} = \frac{\ln 2}{B\epsilon} + O(\epsilon^2) .$$

As an order of magnitude estimate, a computer will fail when the product

$$BC\epsilon = 1 .$$

This result has a simple space-time interpretation, see figure 9. If B devices are spread in space, and the computer executes C time steps, then the total area occupied by the computation in space-time is BC. If any one of those devices in space-time has failed, the computation has failed. That means that $BC\epsilon$ has to be 1 or less.

For workstation class computers the value of $\epsilon$ is very small. The results only apply to the memory of the computer. A typical workstation have (in 1993) around 64 megabytes of memory ($5.4 \times 10^8$ bits). The memory will hold its state for 100 days without an error. Typically memory gets accessed $10^7$ times a second, or $8.6 \times 10^{13}$ times in 100 days. The product BC is then $4.6 \times 10^{22}$, or a failure rate per device (bit) of $\epsilon = 2 \times 10^{-23}$. A very small number.

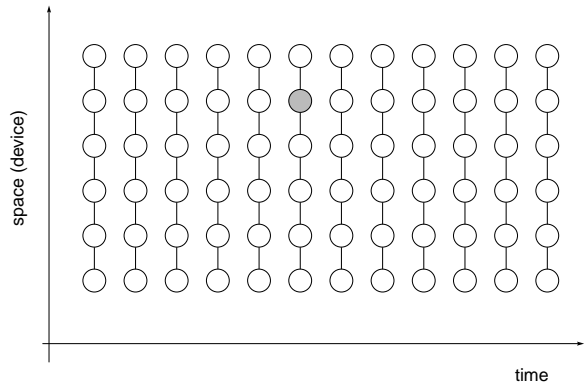

Figure 9: Space-time interpretation of a computation. The computer is made of a series of devices (represented by circles). The state of each device may change as the computation proceeds in time, but if only one device changes state, the whole calculation is wrong.

## 5.2 Error correction

Most nanoscale devices will not have the reliability of the transistor. In some molecular schemes for computation the error rate can approach 10%. With such large errors, a scheme must be found to improve the reliability of computers. Von Neumann was the first to publish on the possibility of building a computer with unreliable parts [24]. He had two motivations: computers were then built with vacuum tubes, which would break often; and to develop a model for the working of real neurons, which can reliably compute despite their variations. The basic idea in von Neumann's work was to increase the reliability of Boolean networks by having them be redundant. If we have an unreliable computer which most of the time produces the right result, but sometimes makes a mistake, then we can have many duplicates of the computer and take a vote of what the correct result should be. The larger the number of computers that are doing the same calculation, the better the chances that the voting will produce the correct result.

The idea of having many copies of the computer is supported by the ideas of Shannon on communication through noisy channels [25]. In Shannon's theory, a message is sent through a noisy communications channel that changes a few bits of the message. Shannon showed that the method to send the message through the channel reliably is to send several copies of the message. Depending on distribution of the errors in the communication channel, the message must duplicate certain bit combinations more than other, but the basic idea continues to be repetition of the message.

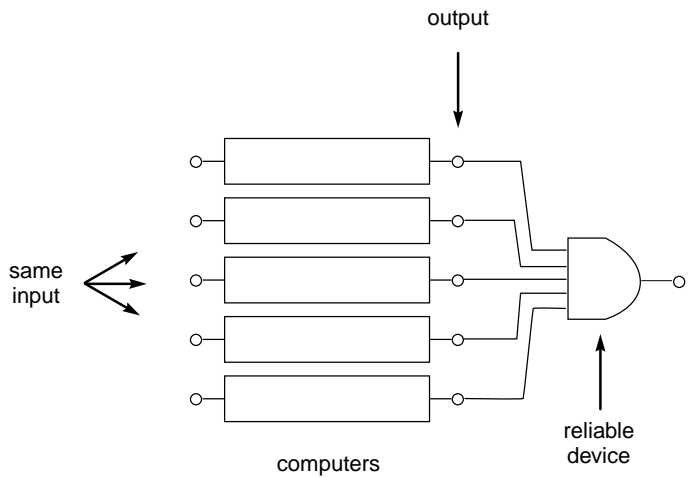

Figure 10: Redundant computation

In my analysis of a general purpose nano-scale computer I will only consider order of magnitude estimates. The working of a computer must be better specified for more detailed estimates. All results will be given as a function of the probability $\epsilon$ that one device has failed. This probability is independent of the state of the computer and on the history of the device. As the results are order of magnitude estimates, I can concentrate on only one error correcting scheme: the majority rule. The majority rule is a voting scheme. Several copies of a device or computer are made to compute or store the same result. The result that occurs more often or in the majority of the cases is considered the state of the device. This is not the most efficient scheme for error correction, but it produces the same order of magnitude (same functional dependence on $\epsilon$) as the more sophisticated error correcting schemes. In particular, the number of times the message is duplicated is the same. This is in agreement with the results of Winograd and Cowan [26]. They have analyzed redundant Boolean networks and used the best information-theoretic coding scheme and obtained the same order of magnitude estimates as von Neumann.

Figure 10 illustrates the majority rule when computing with unreliable components. It also points out a problem in reading the output of error corrected computation. If the computer (rectangle) were reliable, then it would take inputs $i_1, \ldots, i_n$ and produce the output $o_1$. Because it has some chance of making an error one duplicates the inputs and the computer to produce several copies of the result, $o_1, \ldots, o_m$. There is a piece of circuitry that takes the results, decides which occurs more often, and outputs that. This error correcting circuit cannot make errors, as there is no other circuit to check that its results are the majority. This is a general problem with devices that make errors: the repeated output has to be read by a circuit that does not make errors. If it were otherwise, all the error correction done by the duplicated computers would be corrupted by the last output stage. From now on I will assume that the final output can be reliably read.

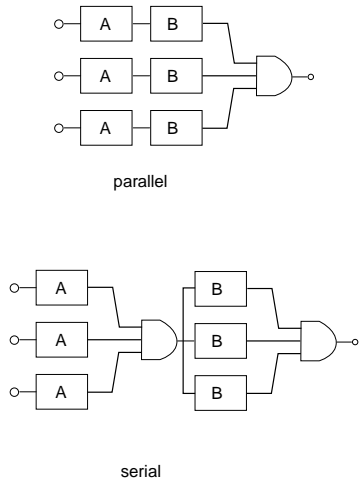

Figure 11: Two possible ways to error correct: in parallel or in series

There are two possibilities when we error correct: we may duplicate the computer several times and error correct at the end; or we may error correct after each part of the computer. To keep the argument simple, break the computer into just two parts, A and B. There are now two possible arrangements, shown in figure 11. In the parallel arrangement the entire computer is duplicated and the error correction is done with the output of each computer. In the serial arrangement the error correction is done with all the stages A in parallel and all the stages B in parallel. Which gives better reliability? To decide this lets use a simple example where the there are only three duplicates of the computer, as shown in figure 11. Assume that the probability that A will fail or that B will fail is $\alpha$ and independent of the failure of other stages. If we have only one computer, as in figure 11(a), the probability that the whole computer will fail is $2\alpha + \alpha^2$. Either one unit fails, and this can happen in two ways; or both units fail.

When the three units are in parallel the probability of the whole computer failing is
$$3(2\alpha + \alpha^2)^2 + (2\alpha + \alpha^2)^3 = 12\alpha^2 + O(\alpha^3) \; ,$$

if $\alpha$ is a small number. When the three units are in series the probability of the whole computer failing is

$$2(3\alpha^2 + \alpha^3) + (3\alpha^2 + \alpha^3)^2 = 6\alpha^2 + O(\alpha^3) \ ,$$

which is of the same order of magnitude as the parallel arrangement. The parallel and the serial scheme are of the same order of magnitude if the probability $\alpha$ of failing is a small number (as compared to the number of stages in the computer). If there are M stages and the product $M\alpha$ is not small, then it is no longer true that the parallel and serial arrangement are of the same order of magnitude. The combinatorial factors overweight the smallness of $\alpha^n$.

When there are many stages, the serial arrangement outperforms the parallel arrangement. Assume that the computer has M stages. In the parallel arrangement the error correction is done by a reliable error correcting device at the end of the M stages. Each stage has a probability $\alpha$ for failing, so the probability $p_f$ that one or more stages will fail in a line and input the wrong result into the error correcting device is

$$p_f = \sum_{k \geq 1} \binom{M}{k} \alpha^k (1-\alpha)^{M-k} = 1 - (1-\alpha)^M \ .$$

If there are $2m-1$ duplicates of each computer, the probability that the whole computer will fail is the probability $p_p$ that more than half of the duplicated computers failed:

$$p_p = \sum_{m \leq k \leq 2m-1} \binom{2m-1}{k} p_f^k (1-p_f)^{2m-1-k} \ .$$

In the serial arrangement there is a reliable error correcting device after a set of $2m-1$ duplicated stages. Each error correcting device outputs the majority input. Again each stage has a probability $\alpha$ of failing. The probability $p_s$ that the whole computer will fail is computed in the reverse order that the failure probability is computed for the parallel arrangement. First we compute the probability that a set of stages will fail even after error correction. That is the probability $p_1$ that m (half) or more stages failed:

$$p_1 = \sum_{m \leq k \leq 2m-1} \binom{2m-1}{k} \alpha^k (1-\alpha)^{2m-1-k} \ ,$$

and the probability the whole computer will fail in the serial arrangement $p_s$ is

$$p_s = \sum_{k \geq 1} \binom{M}{k} p_1^k (1-p_1)^{M-k} = 1 - (1-p_1)^M \ .$$

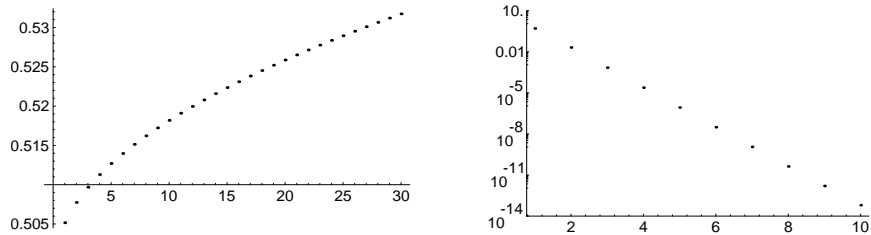

Figure 12: Probability of failure for the whole computer in the parallel and serial arrangements as a function of the number of duplicates m. The computer used in the plots has 70 stages and the probability of failure for any stage, $\alpha$, is 0.01.

If $M\alpha$ is not a small number, there is a large difference between the two arrangements. In figure 12 computers with the same number of stages are compared. In the parallel arrangement the probability starts out growing, whereas in the serial arrangement it rapidly decreases. Notice that the serial plot is logarithmic. The probability of failure $\alpha$ is fixed by the devices used to build a stage of the computer. It is a function of the technology used. The number of stages in the computer is a function of the storage and processing capabilities of the computer. We want these as large as possible. This means that in general the product $M\alpha$ will not be a small number and the serial arrangement will be preferred. It is better to error correct as often as possible.

## 5.3 Correcting with errors

In the previous section we concluded that it is better to error correct after each stage of the computer. The smallest circuit within the computer that could be a stage is the device. It may be a logic gate or storage for a bit. Up to now I have assumed that the error correcting circuit is completely reliable — it introduces no new errors. Reliable error correction will not be possible in a nanocomputer that is integrated to large scales. In a chip, the error correction circuit is built with the same technology as the computer. If few-electron devices are used for the error correcting circuits then it will also be prone to the failures of the logic circuits. Computer circuits and error circuits are all prone to failure by loss of state.

I will now assume a more realistic model for the computer built from unreliable components. Some of the circuit will perform the logic, and some will error correct by the majority rule. The crucial point is that to error correct, signals have to propagate through space, by being carried in wires, or through time, by being stored for a short time. Both forms of propagation increase the chances that the state of a device will get altered. They dramatically change the reliability gains obtained in the serial (and parallel) redundant arrangements (see section 5.2).

I will now consider a unit that replicates the information it is storing. This unit holds a bit while the nanocomputer is carring out a program. After each program cycle the unit stops and executes an error correcting procedure. The unit has $2m - 1$ devices, each of which has to store its state for at least $s$ time steps (cycles) while the error correcting program is executing. Later we will see that $s$ depends on the number of devices in a unit. The probability that any one of the devices will fail is $\epsilon$ and independent of its state and the state of the other devices. This arrangement is the minimal arrangement for error correction. If there are no wires and the nanodevices have to do the error correction, then there have to be a few cycles dedicated to error correction. The program cannot proceed until the error correction part has finished. Also, there is no possibility of having a reliable unit to error correct.

The basic result and main point I wish to make is: the failure rate cannot be made as small as needed with a fixed failure rate per device. This can be seen quite easily. If each device has an failure rate $\epsilon$, and it must hold its state for at least $s$ time steps, then during that period it has a failure rate of $s\epsilon$. The probability that the entire unit will fail is the probability that at least half fail, or $(s\epsilon)^m$. In general $s$ increases as $m^2$, so the probability of the unit failing goes as $(m^2\epsilon)^m$. Asymptotically, this grows with $m$ and the probability cannot be made as small as needed with a fixed $\epsilon$.

The detailed calculation of the probability of failure takes into account the many combinations in which the unit may fail. Unlike the estimate, it is correct for small values of $m$. If a unit with $2m - 1$ devices failed, that means that $m$ or more of the devices did. Each device in the unit has a probability of failure $\epsilon$ independent of its history and the state of other devices in the unit and the computer. Assume that exactly $k$ devices failed. This means that $2m-1-k$ did work for at least one step, and the probability of this happening is $(1-\epsilon)^{2m-1-k}$. The probability that one device worked for at least $s$ steps is $(1-\epsilon)^s$. And the probability that exactly $k$ failed in those $s$ steps is $(1-(1-\epsilon)^s)^k$. The probability that $k$ failed and $2m - 1 - k$ worked is the product of the probability of both events. Combining them, and summing all the ways in which $m$ or more devices in one unit may fail gives the probability $p_u$ that a unit will fail to error correct

$$p_u = \sum_{k \geq m} \binom{2m-1}{k} (1-\epsilon)^{2m-1-k}(1-(1-\epsilon)^s)^k. \qquad (1)$$

For small values of $\epsilon$ this results reduces to the simple estimate.

The number of cycles that each device must keep its state, $s$, is yet to be determined. As we are only interested in an order of magnitude estimate, all we need to determine is how does s depend on the number of copies (redundancy) of each device. More detailed calculations would require a more detailed description of the error correcting procedure. I will give three different estimates for s: one when there are wires available, another when the nanocomputer is two-dimensional, and another when the nanocomputer is one-dimensional. Long wires are equivalent of having many space dimensions: any device can be made neighbor of any others device neighbors by the use of wires. We will find that the lower the dimension in which the computer operates, the harder it is to error correct. In all three cases if the inputs are $x_1, \ldots, x_{2m-1}$, then the majority rule error correcting algorithm has to compute

$$\left\lceil \frac{1}{2m-1} \sum_k (x_k - \frac{1}{2}) \right\rceil ,$$

where $\lceil \cdot \rceil$ returns the smallest integer larger than its argument.

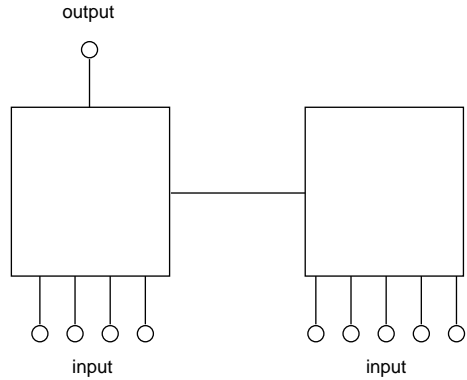

Figure 13: Two halves of a unit trying to decide the majority bit. To avoid fooling the left side output, the right side has to send $\log_2 m$ bits to the left over the wire.

First lets consider the case where wires can be used. Assume that the unit is split into two halves. Each half has some of the devices of the unit (see figure 13). The unit of the left is supposed to determine, after a few cycles, if its state should be the zero bit or the one bit. The unit will determine its state by the majority of the inputs: it will count how many inputs are zero and how many are one. The state occurring most often will be the output. If the unit is to operate successfully, the left unit must know the sum of the one inputs from the right side of the unit. If the right unit does not send a message with at least $\log_2 m$ bits to the left, the left can be fooled into a wrong output. This means that it takes $\log_2 m$ cycles to send the information down a wire from the right part of the unit to the left part of the unit. Keeping the number of devices fixed, this is then a lower bound for the amount of time a unit would take to decide which input is in majority (see reference [27]). A more constructive way of seeing that $\log_2 m$ cycles are needed is to consider an efficient scheme for adding all the input bits without using a lookup table (which would consume too much circuit area), or repeating the inputs (which leads to no gain in efficiency in a nanocomputer). If we use an adder that looks at two inputs and produces an output, then the most efficient adder is arranged in a binary tree. The depth of this tree is $\log_2 m$ and the result of the addition is obtained only after the inputs have advanced from the leaves to the root of the tree. At each cycle one branch level is advanced, so an adder requires $\log_2 m$ cycles to execute. This argument shows that if wires can be used $s$ is at least $\log_2 m$.

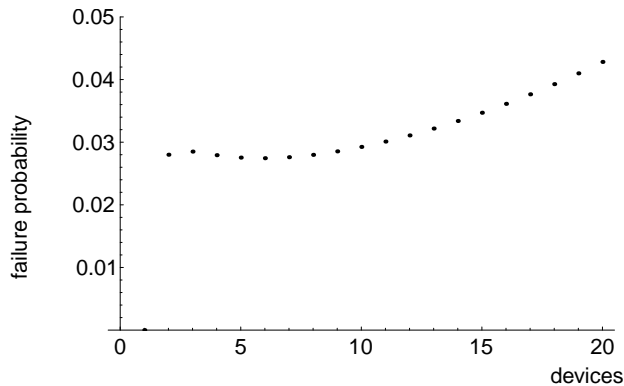

Figure 14: Failure rate of a unit with wires. The number of devices in the unit is $m$ and the probability of failure of a device is $\epsilon = 0.01$.

We can now see how the error rate of a unit behaves when wires are used. In this case $s = \log_2 m$ in the error rate $p_u$ given by equation (1). To illustrate the error probability as the number of devices in a unit increases set $\epsilon = 0.1$ (the probability that a device will fail). The plot is shown in figure 14. The function is very sensitive to the value of $\epsilon$. In general it will have two extrema. For $m = 1$ it is 0, then grows, then decreases to a small value until about $e^{1/(2\epsilon)}$ when it starts growing towards one. For small values of $\epsilon$ (smaller than 0.001 it is possible to reach shallow minima, smaller than $10^{-30}$.

For a nanocomputer each device within a unit must keep its state longer than $\log_2 m$ cycles. This is because there are no wires. Instead the devices are only connected to their nearest neighbors (as in a grid). In executing the algorithm that determines the majority of the states of the unit, there will be a point in the algorithm that one of the devices assumes the majority state.

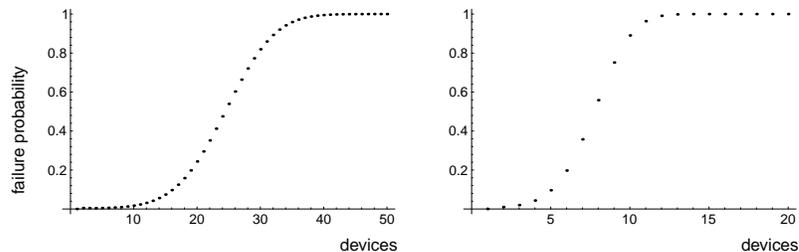

Figure 15: Probability of failure of a unit when there are no wires. For large number of devices $m$ the probability of failure goes to one. The failure probability for a one-dimensional unit is plotted in (a) and for a two-dimensional unit in (b).

To come to that state it had to gather information from all the other devices in the unit. This means that the information on the state of a device has to move of the order $m$ devices. Again we divide the unit into left and right part. The information about the $m$ devices has to be moved across the size of the device, which is of order $\sqrt{m}$ for a two-dimensional unit. This should take $\sqrt{m}$ time steps. As the states of each device are brought to their destination, the number of one bits must be added. That requires $\log_2 m$ cycles for each device. The total time required is then $\sqrt{m} \log_2 m$. If the unit were one-dimensional it would take $m$ time steps to transfer the states from one side to the other of the unit. And the total time to decide the majority bit would be $m \log_2 m$.

In figure 15 I have plotted the probabilities of failure for the one and two-dimensional nanocomputers. The probability is given as a function of the number of devices in one unit, that is, on the redundancy of each bit. The curves have a shallow minimum and then grow to one. The switch from small probability of failure to large probability of failure occurs when $\sqrt{m} \log_2 m = \epsilon^{-1}$ for the two-dimensional case and when $m \log_2 m = \epsilon^{-1}$ for the one-dimensional case. The curves are very sensitive to the value of $\epsilon$ and to the precise dependence of $s$ on $m$. Most practical algorithms for error correction seem to require $s = m^2$. The result to be learned from the curves is that the probability of failure does not go to zero as the number of devices in a unit increases. In figure 16 I have plotted the probability of failure for a unit with three devices. As the probability $\epsilon$ of failure of a device decreases so does the overall probability of failure of a unit decrease.

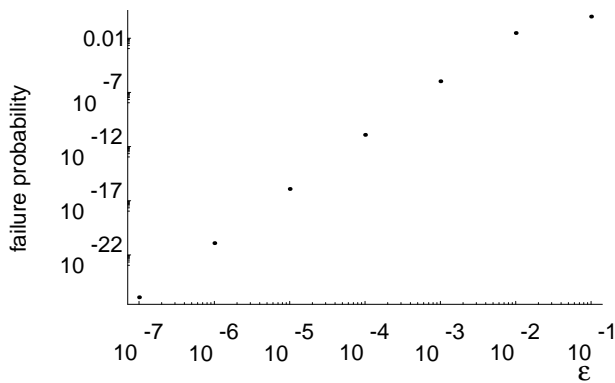

Figure 16: The probability of failure decreases the more reliable each device is. In this example the unit has three devices. Each device has a probability of failure $\epsilon$.

## 5.4 The role of error correction

The plots show that the probability of failure cannot be made as small as needed by replicating the circuits of the computer. It may be that the minimum of probability of failure curve is good enough in practice. But then one has to remember that the calculations only give the dependence of the probability with the number of copies of a device. If we wanted to use a device, despite its unreliable nature, detailed calculations for that device, with all constants, would have to be carried out. If the probability of failure does not go to zero as the number of copies increases, then details matter.

There are two ways to error correct: with and without reliable error correcting devices. These two forms lead to different results for the probability of failure of a unit and of a computer. When the error correction can be done reliably, the error rate of the computer can be made as small as needed. This is done by increasing the number of copies of the computer, gate, or device. When the error correction cannot de done reliably then the crucial point is if there are wires or not in the computer. With wires the information about the many copies of a bit can be quickly transferred to a unit that decides if the bit should be one or zero. The time it takes to error correct is almost independent of the number of copies (depends on the logarithm). Without wires the time it takes to error correct increases with the number of copies. If the number of copies is large, it takes too long to error correct, and the error correction procedure is itself error prone.

Error correction also decreases the speed of a computer. For each cycle of computation, several cycles must be used to error correct. In the case of a nanocomputer with units that error correct by using $m$ copies, the fraction of time not error correcting is $1/\sqrt{m}$. The error correction cannot be done by just replicating the whole computer many times and using reliable error correcting devices. This is the parallel arrangement which has errors that do not go to zero. What is needed is a figure of merit that takes into account the tradeoffs between the area occupied by the computer, the time it takes to perform a calculation, and the failure probability.

Other aspects of error in computers have not been analyzed. When very thin wires are used, or when the devices are small, the connections among them may fail. Connections may also be wrong. The yield of a complicated circuit may be too small if all connections have to be correct. As von Neumann showed [24], it is also possible to error correct for wrong connections by replicating the circuits of a computer.

## 6 Stochastic computation

The prospects for computation with nano-scale devices seem gloomy. Finding a device that has enough gain to overcome the errors may not be possible. But this does not mean that computation, useful computation, is not possible with nano-scale devices. What has to change is the manner in which computation is done. We can give up the notion of a general purpose computer and have the nano-computer execute the expensive part of the calculation. For example, a fast nano-scale computer could be coupled to a slower general purpose computer. The slower computer could set up the calculation for the faster computer, taking care of input and output. The faster computer would then execute the expensive and specialized part of the computation.

In principle, the special nano-computer could be used to simulate a general purpose computer. If it is fast enough (as nano-scale electronics seem to promise) the extra level of simulation will not be an obstacle for an useful computer. There is a problem with this argument. Because the nano-computer will have to use active wires to transmit information from one end of the circuit to another, many of the clock cycles will be would be used for information propagation. The time is of the order of the linear size of the circuit, or $\sqrt{m}$ with $m$ the number of gates. With typical circuits having of the order of $10^6$ gates in 1994, a speed-up of at least $10^3$ would be needed, taking the typical operating frequency from 50 MHz to 50 GHz. This is a pessimistic estimate, but it does point to the dangers of digital error correction.

All the estimates and arguments on the difficulties of using nano-scale computers assume that traditional von Neumann style algorithms will be executed on the computer. Many of the arguments *may* not apply if other algorithms are used. As an illustration of the algorithms I have in mind, I will develop the idea of stochastic computation. Stochastic computation is a simple lattice gas [28]. The information is stored as the number of bits in a region of the computer. The results of the computation are obtained by averaging large regions containing many electrons. The averaging, and the analog nature of the result, make the system immune to small errors.

I will take as a model for a computer the simplest possible device that can be nanofabricated: a large array of quantum dots. In this array each dot is connected to its neighbors by tunneling junctions. Such arrays have not yet been fabricated, but each of the repeating elements has already been built. Array of dots without interconnections have been built and can be charged in a controlled way with a small number of electrons [29]. The array is very simple, with only three elements per repeating unit. If more sophisticated devices can be built, the ideas presented on computation could be more easily applied. First I will show how an array of quantum dots can solve a simple problem — the diffusion equation — and later how it can be generalized to other partial differential equations such as the Schrödinger equation.

Let us abstract the array of dots as a lattice of sites that connect to their nearest neighbors. Each site can assume a finite number of states. The question is: "can this array be used for computation?" One simple application is to the solution of the diffusion equation:

$$\partial_t \rho = D \nabla^2 \rho.$$

The solution for this equation at a time T can be found from an initial condition $\rho(0)$ by starting a series of non-interacting random walkers on the lattice, distributed so as to approximate the initial distribution $\rho(0)$. The motion of the random walkers consists of hopping to a neighboring lattice site at each time step. All neighboring sites are equally probable. The random walkers can be made to solve a slightly more complicated equation:

$$\partial_t \rho = D \nabla^2 \rho - V(x) \rho.$$

The solution to this equation is found by using random walkers that jump to the different neighboring sites with probabilities proportional to the potential at the neighboring site.

A simple physical realization of a stochastic computer solving the diffusion equation is the array in figure 17. The circles represent quantum dots that can hold a few electron states, and the rectangles are single electron tunneling devices. The dots represent the lattice sites in the diffusion problem, and the tunneling junctions control the hopping probabilities from one site to its neighbors. This is a computer that could be built from conductors and insulators, and not just semiconductors, see Likaharev [30]. Above each tunneling junction there is a conducting plate that controls the tunneling rate of the junction. The field ρ of the equation is obtained by averaging over a large number of quantum dots.

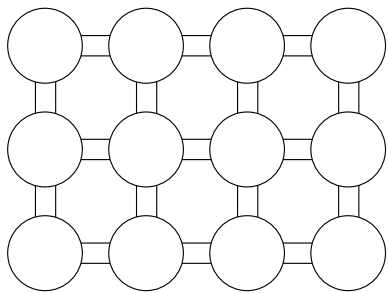

Figure 17: A basic computer. The circles are quantum dots and the rectangles are tunneling junctions. With this simple computer it is possible to solve partial differential equations.

The metal-insulator construction is not the only possible one. More conventional technology could be used. What is required for a stochastic computer is a storage device that can hold a few bits and a logic device that reads the bits. The logic device should be capable of reading the the bits from two neighboring sites, execute a simple logical operation with them, and then store the result in a third site. For the storage devices one could use multiple state resonant tunneling diodes [31, 32]. The logic device could be fabricated with lateral resonant tunneling devices [33, 34, 35]. Similar types of computers have been proposed before, see Biafore [36] and Lent *et al.* [37].

The elements of the stochastic computer do not have to be quantum dots and tunneling junctions. More conventional circuit elements could be used.

A computer built to solve the diffusion equation would not be of much practical value. Its importance comes when we realize that what is actually being computed is a Wiener process — an Euclidean path integral [38]. Most of the problems of modern theoretical physics fall in this class. For example, the stationary Schrödinger equation can be transformed into a diffusion problem by introducing a fictitious time. Other examples include the Poisson equation, the Fokker-Plank equation, and the Navier-Stokes equation.

Let us say that we wanted to determine the ground state of the Schrödinger equation subject to a field $v(x)$. For the ground state and a real external field it always possible to arrange the wave function $\psi(x)$ to be real and positive. Take

the Schrödinger equation

$$-\frac{1}{2}\nabla^2\psi(x) + v(x)\psi(x) = \epsilon\psi(x),$$

and substitute

$$\rho(x,\tau) = \psi(x)e^{-\epsilon\tau},$$

where $\tau$ is the fictitious time. If we now compute $\partial_t\rho$ one finds that

$$\partial_\tau\rho = \frac{1}{2}\nabla^2\rho - v(x)\rho$$

which is a diffusion equation with an external field. To, determine the eigenenergy $\epsilon$ one just has to analyze the rate of decay of the solution.

The example is not limited to the ground state of the Schrödinger equation. By having four states per site it is possible to determine wave functions with nodes. It is also possible to apply the diffusion method to other equations [39].

The advantage of having the equation solved by a Wiener process is that the irregularities of the array can be used to ones advantage in simulating the process. Before a given process is simulated the array can be calibrated with a simple diffusion process. An initial distribution of electrons would be injected into the array and let to evolve. It would then be compared to the exact solution. The difference between the simulation and the exact solution would give amount of bias that should be applied to each of the tunneling regions so that the simulation and solution agree. Then when a different process has to be simulated its external field would be added to the bias. This procedure could correct for the imperfections in the manufacture of the array.

## 7 Conclusions

The field of nanoelectronics is a rapidly changing field. New devices are constantly being created. To avoid technology forecasting [40] I have identified the basic properties of nanodevices and how they can be used for computing. The long argument explained why etched wires cannot be used between all devices of the computer (they open up due to electromigration), and why digital error correction cannot improve the failure rate (quantum devices make errors while error correcting). These results are intuitive to those that build devices. I have tried to explain how they follow from the basic principles used to design computers and from the basic properties of quantum devices.

The results on long wires and error correction are order of magnitude estimates. Error rates cannot be made arbitrarily small for a fixed error rate per device. It may be that a given device, when used in a very large scale integrated circuit, will have acceptable error rates. It may be that this circuit can be operated slow enough to avoid electromigration. The exact figures can only be found by building it.

Many of these results do not apply to devices under research. For example, if a wire can be grown (self-assembled) as a single crystal grain, its lifetime is extended by several orders of magnitude. Also, if a process is found for the manufacturing of integrated circuits that guarantees, from physical principles, that all devices are dimensioned to the atomic scale, then the amount of gain needed is reduced. Even if these new manufacturing processes are not found, stochastic computers can still be built with current technology. The problem then reduces to finding a large enough class of algorithms to justify their development and use.

All the devices needed for the fabrication of the stochastic computer at the nanoscale have been demonstrated. The storage devices can be built out combinations of resonant tunneling diodes. The logic units can be built out of dual gate resonant tunneling field effect logic arrays. If this computer could operate without any errors, it would be equivalent to a Turing machine. There are at least two very useful algorithms to execute in this computer: the diffusion equation (which computes the ground state of the Schrödinger equation) and the FHP gas [28] (which solves the Navier-Stokes equation).

## Acknowledgments

I would like to thank Michael Biafore, Robert Ecke, Brosl Hasslacher, and Sid Singer for useful discussions. I would also like to thank Mark Reed for allowing me to adapt one of his slides for this publication (figure 5). This work was supported by the Department of Energy.

## References

[1] Intel factory to grow at cost of $1 billion. Wall Street Journal, Friday, 2 April 1993.

[2] F. Dyson. *Infinite in all directions*. Harper & Row, New York, 1988. See the chapter "Quick is beautiful".

[3] Wojciech Zurek, editor. *Complexity, entropy, and the physics of information*, volume 8 of *Proceedings volume in the Santa Fe Institute studies in the sciences of complexity*. Addison-Wesley, Redwood City, 1990.

[4] Robert W. Keyes. What makes a good computer device? *Science*, 230:138–144, 1985.

[5] Tommaso Toffoli and Norman Margolus. *Cellular automata machine: a new environment for modeling*. MIT Press, Cambridge, 1987.

[6] Bernard S. Landman and Roy L. Russo. On pin versus block relationship for partitions of logic graphs. *IEEE Transactions on computers*, 20:1469–1479, 1971. This article describes the experiments that Rent carried out at IBM.

[7] David K. Ferry. Interconnection lengths and VLSI. *IEEE Circuits and Devices Magazine*, 1(4):39–42, July 1985.

[8] A. Mukhopadhyay. Hardware algorithms for nonnumeric computation. *IEEE Transactions on computers*, 28(6):384–394, 1979.

[9] R. W. Floyd and J. D. Ullman. The compilation of regular expressions into circuits. *Journal of the A.C.M.*, 29(2):603–622, 1982.

[10] Jeffery D. Ullman. *Computational Aspects of VLSI*. Computer Science Press, Rockville, 1984.

[11] Claude E. Shannon. The synthesis of two-terminal switching circuits. *Bell Systems technical Journal*, 28:59–98, 1949.

[12] B. McColl. Planar circuits have short specifications. In *Proceedings of the Second STACS 1985*, volume 182 of *Lecture Notes in Computer Science*, pages 231–242. Springer-Verlag, 1985.

[13] Carver Mead and Lynn Conway. *Introduction to VLSI systems*. Addison-Wesley, Reading, 1980.

[14] R. Landauer. Spatial variation of currents and fields due to localized scatters in metallic conduction. *IBM Journal of Research and Development*, 1:223–231, 1957.

[15] M. Büttiker. Four terminal phase coherent conductance. *Physical Review Letters*, 57(14):1761–1764, 1986.

[16] M. L. Roukes and O. L. Alerhand. Mesoscopic junctions, random scattering, and strange repellers. *Physical Review Letters*, 65(13):1651–1654, 1990.

[17] Ajit Ram Verma and P. Krishna. *Polymorphism and polytypism in crystals*. Wiley, New York, 1966.

[18] Thomas Kwok. Electromigration and reliability in submicron metallization and multilevel interconnections. *Materials Chemistry and Physics*, 33:176–188, 1993.

[19] Gunther Bauer, Friedemar Kucher, and Helmut Heinrich, editors. *Low-dimensional electronic systems: new concepts.*, volume 111 of *Springer series in solid-state sciences*, Berlin, 1992. Springer-Verlag.

[20] Leo Esaki. New phenomenon in narrow Germanium *p-n* junctions. *The Physical Review*, 109:603–605, 1958.

[21] L.L. Chang, L. Esaki, and R. Tsu. Resonant tunneling in semiconductor double barriers. *Applied Physics Letters*, 24:593–595, 1974.

[22] Mark A. Reed and Wiley P. Kirk, editors. *Nanostructure physics and fabrication: proceedings of the international symposium*, Boston, 1989. Academic Press.

[23] Anant G. Sabnis. *VLSI Reliability*, volume 22 of *VLSI Eletronics Microstructure Science*. Academic Press, San Diego, 1990.

[24] J. von Neumann. Lectures. In C. E. Shannon and J. McCarthy, editors, *Automata Studies*. Princeton University Press, 1956.

[25] C. E. Shannon. A mathematical theory of communication. *Bell System Technical Journal*, 27:379–423, 623–656, 1948.

[26] S. Winograd and J. D. Cowan. *Reliable computation in the presence of noise*. M.I.T. Press, Cambridge, 1963.

[27] Thomas Lengauer. VLSI theory. In J. van Leeuwen, editor, *Handbook of theoretical computer science*, volume A, chapter 16, pages 835–868. Elsevier Science, Amsterdam, 1990.

[28] U. Frisch, B. Hasslacher, and Y. Pomeau. Lattice-gas automata for the Navier-Stokes equation. *Physical Review Letters*, 56(14):1505–1508, 1986.

[29] B. Meurer, D. Heitmann, and K. Ploog. Single electron charging of quantum dot atoms. *Physical Review Letters*, 68(9):1371–1374, 1992.

[30] K. K. Likharev. Correlated discrete transfer of single electrons in ultrasmall tunnel junctions. *IBM Journal of Research and Development*, 32:144–158, 1988.

[31] A. Zaslavsky, V. J. Goldman, D. C. Tsui, and J. E. Cunningham. Resonant tunneling and intrinsic bistability in asymmetric double barrier heterostructures. *Applied Physics Letters*, 53:1408–1410, 1988.

[32] Harold J. Levy and T. C. McGill. A feedforward artificial neural network based quantum effect vector matrix multiplier. *IEEE Transactions on neural networks*, 4:427–433, 1993.

[33] J. N. Randall, A. C. Seabaugh, and J. H. Luscombe. Fabrication of lateral resonant tuneling devices. *Journal of Vacuum Science and Technology B*, 10(6):2941–2944, 1992.

[34] S. Y. Chou, D. R. Allee, R. F. W. Pease, and J. S. Harris Jr. Observation of electron resonant tunneling in a lateral dual-gate resonant tunneling field-effect transistor. *Applied Physics Letters*, 55:176–178, 176.

[35] Susanata Sen, Federico Capasso, Alfred Y. Cho, and Debbie Sivco. Resonant tunneling device with multiple negative differential resistance: digital and signal processing applications with reduced circuit complexity. *IEEE Transactions on electronic devices*, 34:2185–2191, 1987.

[36] Michael Biafore. Cellular automata for nanometer-scale computation. *Physica D*, 70(4):415–433, 1994.

[37] Craig Lent, P. Douglas Tougaw, Wolfgang Porod, and Gary H. Bernstein. Quantum cellular automata. *Nanotechnology*, 4:49–57, 1993.

[38] Claude Itzykson and Jean-Michel Drouffe. *Statistical field theory*. Cambridge University Press, Cambridge, 1989.

[39] R. Marra. Probabilistic approach to the Navier-Stokes equation. *Physics Letters A*, 148:41–44, 1990.

[40] Rolf Landauer. Advanced technology and truth in advertising. *Physica A*, 168:75–87, 1990.